\documentclass[12pt]{article}
\usepackage{amsfonts}
\usepackage{amsmath}
\usepackage{graphicx}
 \usepackage{epstopdf}
\newcommand{\be}{\begin{equation}}
\newcommand{\ba}{\begin{eqnarray}}
\newcommand{\ee}{\end{equation}}
\newcommand{\ea}{\end{eqnarray}}
\newcommand{\cosech} { {\rm cosech}}
\newcommand{\sech} { {\rm sech}}
\DeclareMathOperator{\erf}{erf}

\begin{document}

\title{ One parameter family of rationally extended isospectral potentials}

\author{Rajesh Kumar Yadav$^{a}$\footnote{e-mail address: rajeshastrophysics@gmail.com (R.K.Y)}, Suman Banerjee$^{a}$\footnote{e-mail address: suman.raghunathpur@gmail.com (S.B)}, Nisha Kumari$^{b}$\footnote{e-mail address: nishaism0086@gmail.com (N.K)}, Avinash Khare$^{c}$\footnote {e-mail address: khare@physics.univpune.ac.in (A.K)},  and \\
  Bhabani Prasad Mandal$^{d}$\footnote{e-mail address: bhabani.mandal@gmail.com (B.P.M).}}
 \maketitle
{$~^a$Department of Physics, S. K. M. University, Dumka-814110, INDIA.\\
$~^b$Department of Physics, S. P. College, Dumka-814101, INDIA.\\
$~^c$Department of Physics, Savitribai Phule Pune University, Pune-411007, INDIA.\\
$~^d$Department of Physics, Banaras Hindu University, Varanasi-221005, INDIA. }

\begin{abstract}
We start from a given one dimensional rationally extended potential 
associated with $X_m$ exceptional orthogonal 
polynomials and using the idea of supersymmetry in quantum 
mechanics, we obtain one continuous parameter ($\lambda$) family of 
rationally extended strictly isospectral potentials whose solutions are also 
associated with $X_m$ exceptional orthogonal polynomials. We illustrate this 
construction by 
considering three well known rationally extended potentials, two with pure discrete spectrum 
(the extended radial oscillator and the extended Scarf-I) and one with both 
the discrete and the continuous spectrum  
(the extended generalized P\"oschl-Teller) and explicitly construct the 
corresponding one continuous parameter family of rationally extended strictly 
isospectral potentials. Further, in the special case 
of $\lambda=0$ and $-1$, we obtain two new exactly solvable rationally 
extended  potentials, namely the  
rationally extended 
Pursey and the rationally extended Abhrahm-Moses potentials respectively. 
We illustrate the whole procedure by discussing in detail the 
particular case 
of the $X_1$ rationally extended one parameter family of potentials including
the corresponding Pursey and the Abraham Moses potentials.

\end{abstract}
\section{Introduction}
Since the advent of supersymmetric (SUSY) quantum mechanics \cite{cks,ks} 
there has been 
a flood of activity in discovering new exactly solvable potentials. Amongst
these, special mention may  be made of isospectral potentials which have 
application in several diverse areas like
in the problems of inverse scattering \cite{invs1,invs2}, $\alpha\alpha$- 
scattering \cite{baye}, soliton theory \cite{sol1,sol2} etc. This has 
motivated the researchers to search for a family of isospectral potentials 
\cite{nieto,amado,pursey,cv,amp}. Several different approaches have been used
for this purpose such as Darboux method \cite{darboux}, 
Abraham and Moses method \cite{amp}, Pursey method \cite{pursey}, and 
Supersymmetric (SUSY) method \cite{akus} etc to discover new isospectral 
potentials. 
In a very interesting paper Pursey \cite{pursey} has  
shown that the first three methods are inequivalent in the case of inserting 
or deleting states while for normalizing states these are equivalent  
to that of Abraham and Moses approach (which is based on Gelfand-Levitan 
approach \cite{gele}). 
Subsequently, Amado \cite{amado} followed the Pursey approach and constructed 
the potentials isospectral to the coulomb potential with one less bound state.
Out of all these approaches, the SUSY approach is simpler and convenient in 
searching for a family of isospectral potentials.
In particular, using this approach it was shown \cite{nieto,akus} 
that for any given 
central or one dimensional potential with at least one bound state, a 
continuous one-parameter ($\lambda$)
family of strictly isospectral potentials can be obtained. They further showed 
that as $\lambda\rightarrow 0$, the potential reduces to the 
Pursey potential \cite{pursey} and as $\lambda\rightarrow -1$, it reduces to 
the Abraham-Moses [AM] potential \cite{amp}. 
It has also been shown that the partner potential has the 
same bound state spectrum as those of the Pursey and the AM potentials. They 
considered the usual Coulomb and the radial oscillator
potentials and discussed the various isospectral potentials following from 
them. 

Recently, after the discovery of the $X_m$-exceptional orthogonal polynomials
(EOPs) ($m\geq 1$) 
\cite{eopm1,eopm2,eopm3}, many of the well known shape invariant potentials 
have been 
 extended rationally and their 
exact solutions have been obtained in terms of these EOPs 
\cite{que,bqr,os,hos,hs,que8,op1,op2,op3,midya,rkyd}. Using supersymmetric 
(SUSY) quantum 
mechanics \cite{cks,ks}, it has been shown that these rationally extended (RE) potentials are also 
shape invariant. Various properties of these extended potentials have been 
 studied in \cite{midyapd, midya1,clh11,dr11,scatt1,scatt,scatt4,n16,nk16,
ramos, para, nk17, nk18, bbp,rkmany}.
It is worth pointing out that the corresponding (non-rational) potentials are 
the special $m=0$ case 
 of the related RE potentials. 

One interesting question is how to extend the number of analytically solvable
RE potentials. In this context, we observe that 
to the best of our knowledge  
so far one continuous parameter family of strictly isospectral potentials 
(i.e. those with identical discrete spectrum and identical reflection and
transmission amplitudes in the one dimensional case or identical scattering 
amplitude in the three 
dimensional case) have 
only been obtained for
 the case of the conventional (non-rational) potentials, but no such 
construction exists for the RE potentials. Similarly, while Pursey and
AM potentials 
corresponding to the 
usual (non-rational) potentials are known, the corresponding rational 
Pursey and AM potentials are still not available.  
The purpose of this paper is to fill this gap and in this process also extend
the number of analytically solvable isospectral RE potentials. 

In this paper, we consider RE potentials 
and follow the formalism of SUSY quantum mechanics and generate 
one continuous parameter
($\lambda$) family of exactly solvable RE strictly isospectral potentials. 
In the
limit of $\lambda =0$  and $\lambda =-1$, we then obtain the corresponding 
RE Pursey and the RE AM potentials respectively. 
We elaborate this construction by considering three well known RE
potentials, two with pure discrete spectrum 
(i.e. the extended radial oscillator and the Scarf-I potentials) and one 
with discrete as well as the continuous spectrum (i.e. the extended 
generalized P\"oschl-Teller potential) and for all three cases construct 
the corresponding one parameter 
family of strictly isospectral  
RE potentials. In other words, if we start from these three conventional (non-rational) potentials then
one has constructed two parameter family of strictly isospectral potentials. Out of these two, $\lambda$ is a continuous parameter while
$m$ is a discrete parameter ($m = 1, 2, 3,...$) corresponding to $X_m$ EOPs. 
Further, we also obtain the corresponding
one discrete parameter $m$ family of RE Pursey and RE AM potentials. 

The organization of this paper is as follows: In Sec. $2$ we briefly discuss
the SUSY formalism and explain how by starting from any potential with at least one bound state one can construct one continuous parameter ($\lambda$)
family of strictly isospectral potentials. 
Further we also point out how to 
obtain the corresponding Pursey and the AM potentials and explain how
their eigenfunctios are related to those of the original potential with 
at least one bound state.  
In Sec. $3$, 
we consider three RE potentials associated with the $X_m$ EOPs and construct 
the corresponding one continuous parameter ($\lambda$) 
family of the RE strictly isospectral potentials $\hat{V}^{(-)}_m(x, \lambda)$ with their solutions in terms of 
$X_m$ EOPs. For illustration, we consider the $X_1$ ($m=1$) case 
in detail for all three RE potentials
and give plots of $\hat{V}^{(-)}_1(x, \lambda)$ for several values of $\lambda$ as well 
as the corresponding ground state
eigenfunctions for all three RE potentials. Further, we also give plots of the
corresponding three isospectral potentials $V^{[+]}_1, V^{[P]}_1$ and $V^{[AM]}_1$. 
Finally, we summarize our results 
in Sec. $4$.

 \section{Supersymmetric (SUSY) Quantum Mechanics formalism}

In this section, we briefly review the SUSY QM formalism \cite{akus} to 
construct  
 one continuous parameter ($\lambda$) family of strictly isospectral 
potentials corresponding to any potential with at least one bound state.
Given a Hamiltonian with ground state energy $E_0$, one can trivially 
construct a new Hamiltonian from here with zero ground state energy. In the 
SQM formalism one considers a Hamiltonian ($\hbar = 2m = 1$) 
\be
H^{(-)}(x)=-\frac{d^2}{dx^2}+V^{(-)}(x), 
\ee
 with zero ground state energy and the corresponding ground state eigenfunction
being $\psi^{(-)}_{0}(x)$. It can be factorized in terms 
of operators $A$ and $A^{\dagger}$ as 
\be
H^{(-)}(x)=A^{\dagger}A 
\ee 
with
\be
A=\frac{d}{dx}+W(x) \quad \mbox{and} \quad A^{\dagger}=-\frac{d}{dx}+W(x),
\ee
where 
\be\label{sup}
W(x) =-\frac{d}{dx}[\ln\psi^{(-)}_{0}(x)]
\ee 
is the superpotential, which determines the two partner potentials 
\be\label{pp}
V^{(\pm)} (x)=W^2(x)\pm W'(x).
\ee
The eigenvalues
and the eigenfunctions of these two partner potentials 
(when the SUSY is unbroken) are related by 
\be\label{ev}
E^{(-)}_{n+1}=E^{(+)}_n \qquad  E^{(-)}_{0}=0,
\ee
and 
\be\label{pwf}
\psi^{(+)}_{n}(x)=\frac{1}{[E^{(+)}_n]^{1/2}}A\psi^{(-)}_{n+1} \qquad \psi^{(-)}_{n+1}(x)=\frac{1}{[E^{(+)}_n]^{1/2}}A^{\dagger}\psi^{(+)}_{n}
\ee
respectively. For the one dimensional case, the transmission $(t^{(\pm)}(k))$ and reflection $(r^{(\pm)}(k))$ amplitudes 
for the partner potentials $V^{(\pm)}(x)$ are related by
\be\label{t}
r^{(-)}(k)=\bigg(\frac{W_{-}+ik}{W_{-}-ik}\bigg)r^{(+)}(k)
\ee
and 
\be\label{r}
t^{(-)}(k)=\bigg(\frac{W_{+}-ik'}{W_{-}-ik}\bigg)t^{(+)}(k)
\ee
where 
\be\label{kk}
k=(E -W^2_{-})^{\frac{1}{2}}\quad \mbox{and}\quad k'=(E -W^2_{+})^{\frac{1}{2}}
\ee
with      
\be\label{scatt}
W_{\pm} =W(x\rightarrow \pm \infty).
\ee
Similarly, for $3D$-central potentials the scattering amplitudes for the partner potentials are related by \cite{cks}
\be\label{sa}
s^{(-)}(k')=\bigg(\frac{W_{+}-ik'}{W_{+}+ik'}\bigg) s^{(+)}(k').
\ee
\subsection{Generation of isospectral potentials}
The one-parameter family of potentials $\hat{V}^{(-)}(\lambda,x)$ which are 
strictly isospectral to a given potential $V^{(-)}(x)$ 
are obtained by considering the uniqueness of the partner potential 
$V^{(+)}(x)$ \cite{akus}, i.e. for a given more general 
form of the superpotential $\hat{W}(x)$, $V^{+}(x)$  satisfies 
\be\label{gpat}
V^{(+)}(x)=\hat{W}^2(x) +\hat{W}'(x) = W^2(x) + W'(x)\,. 
\ee
Here $\hat{W}(x)=W(x)$ is one of the solution of the above equation. The most 
general solution to the Eq. (\ref{gpat}) can be obtained 
by defining 
\be\label{gs}
\hat{W}(x)=W(x)+\phi(x).
\ee 
On using (\ref{gs}) in (\ref{gpat}) and compare with (\ref{pp}), we get
\be\label{req}
\phi'(x)+2W(x)\phi(x)+\phi'(x)=0.
\ee 
If we set $y(x)=\frac{1}{\phi(x)}$, we get the Bernoulli equation 
\be
y'(x)=1+2W(x)y(x).
\ee
with the solution 
\be
\frac{1}{y(x)}=\phi(x)=\frac{d}{dx}\ln[I(x)+\lambda],
\ee
where the integral $I(x)$, in term of the normalized ground state wavefunction, 
is given by
\be\label{intim}
I(x)=\int^{x}_{-\infty}[\psi^{(-)}_{0}(x)]^2dx
\ee
and $\lambda$ is an abitrary real constant. Thus the most general form of the 
superpotential $\hat{W}(x)$ satisfying Eq. (\ref{gpat}) 
is given by 
\be\label{spiso}
\hat{W}(x)=W(x)+\frac{d}{dx}\ln[I(x)+\lambda]\,.
\ee 
Thus one obtains one continuous parameter family of potentials 
\be\label{isop}
\hat{V}^{(-)}(\lambda,x)=\hat{W}^2(x)-\hat{W}'(x)=V^{(-)}(x)
-2\frac{d^2}{dx^2}\ln(I(x)+\lambda)\,,
\ee   
all of which have the same SUSY partner potential $V^{(+)}(x)$.
From Eqs. (\ref{sup}) and (\ref{spiso}), the associated normalized ground 
state eigenfunction for the potential $\hat{V}^{(-)}(\lambda,x)$ is  
given by 
\be\label{gswf}
\hat{\psi}^{(-)}_{0}(\lambda,x)=\frac{\sqrt{\lambda (1+\lambda )}\psi ^{(-)}_{0}(x)}{[I(x)+\lambda ]},
\ee
and from Eqs. (\ref{pwf}) and (\ref{spiso}) the normalized excited-state ($n=0,1,2,...$) eigenfunctions are
\be\label{extwf}
\hat{\psi}^{(-)}_{n+1}(\lambda,x)=\psi^{(-)}_{n+1}(x)+\frac{1}{E^{(-)}_{n+1}}\bigg(\frac{I'(x)}{I(x)+\lambda}\bigg)\bigg( \frac{d}{dx}+W(x)\bigg)\psi^{(-)}_{n+1}(x).
\ee
The associated energy eigenvalues are
\be
\hat{E}^{(-)}_{n+1}=E^{(+)}_{n}\,,~~E^{(-)}_{0} = 0\,,~~n = 1, 2,...\,..
\ee

From Eqs. (\ref{intim}) and (\ref{gswf}) it is clear that the ground state
eigenfunction is acceptable only if $\lambda > 0$ or $\lambda < -1$ and in
these cases the entire one continuous parameter family of potentials have
the same energy eigenvalues as $V^{(-)}(x)$. Besides, in case $\lambda > 0$
or $\lambda < -1$  
\be\label{2.2}
\hat{W}(x\rightarrow \pm\infty)= W(x \rightarrow \pm \infty) = {W}_{\pm}\,,
\ee 
and hence in these cases the scattering amplitudes 
are also the same as those of $V^{(-)}(x)$ i.e. are given by Eqs. (\ref{t}) and 
(\ref{r}) in the one dimensional case or by Eq. (\ref{scatt}) in the three
dimensional case. Summarizing, given any potential $V^{(-)}(x)$ with at least one
bound state, one can easily construct one continuous parameter family of
potentials $V^{(-)}(x, \lambda)$ which are strictly isospectral to the potential 
$V^{(-)}(x)$ provided $\lambda > 0$ or $\lambda < -1$.

On the other hand when $\lambda=0$ or $\lambda=-1$, then   
$I(x)+\lambda$ with $I(x)$ given by Eq. (\ref{intim}) will vanish either at 
$x\rightarrow -\infty$ or $x\rightarrow +\infty$ respectively, so that the 
ground state eigenfunction 
$\hat{\psi}^{(-)}_{0}(\lambda,x)$ as given by Eq. (\ref{gswf}) is no more 
square integrable and hence the SUSY is broken so that the bound state 
spectrum is degenerate 
with the partner potential $V^{(+)}(x)$, i.e. in this case the potential 
$\hat{V}^{(-)}(\lambda,x)$ loses a bound state. The potentials 
corresponding to $\lambda=0$ and $-1$ are the so called Pursey $V^{[P]}(x)$ 
and the Abraham-Moses 
$V^{[AM]}(x)$ potentials respectively \cite{akus}. For completeness we now
mention 
some of the important results of these two potentials. 
  
\subsubsection{The Pursey potential}
 
The superpotential for this case is defined by putting $\lambda=0$ in 
Eq. (\ref{isop})    
\be\label{ps}
W^{[P]}(x)=W(x)+\frac{d}{dx}\ln I(x).
\ee
and the potential (\ref{isop}) becomes
\be\label{pur}
V^{[P]}(x)=\hat{V}^{(-)}(\lambda=0,x)=V^{(-)}(x)-2\frac{d^2}{dx^2}\ln I(x).
\ee
Since the supersymmetry between $V^{(+)}(x)$ and $V^{[P]}(x)$ is broken, thus 
the energy eigenvalues of this potential are isospectral to those of $V^{(+)}(x)$,
i.e. 
\be
E^{[P]}_n=E^{(+)}_n.
\ee 
Further, the scattering amplitudes for one dimensional case are 
\be\label{ref}
r^{[p]}(k)=\bigg(\frac{W_{-}-ik}{W_{-}+ik}\bigg)^2 r^{(-)}(k)
\ee
\be\label{tamp}
t^{[p]}(k)=-\bigg(\frac{W_{-}-ik}{W_{-}+ik}\bigg)t^{(-)}(k)
\ee
while for the three dimensional case 
\be\label{scattp}
s^{[P]}(k')=s^{(-)}(k').
\ee
The normalized eigenfunctions $\psi^{[P]}_{n}(x)$ for the Pursey potentual 
$V^{[P]}(x)$ are 
obtained by putting $\lambda=0$ in Eq. (\ref{extwf}) i.e.,
\ba\label{extwfp}
\psi^{[p]}_{n}(x)&=&\hat{\psi}^{(-)}_{n+1}(\lambda=0,x)\nonumber\\
&=&\psi^{(-)}_{n+1}(x)+\frac{1}{E^{(-)}_{n+1}}\bigg(\frac{I'(x)}{I(x)}\bigg)\bigg( \frac{d}{dx}+W(x)\bigg)\psi^{(-)}_{n+1}(x),
\ea
where $n=0,1,2...,$.
\subsubsection{The Abraham-Moses potential} 
In this case since $\lambda = -1$, the superpotential and the potential 
are given by 
\be\label{psup}
W^{[AM]}(x)=W(x)+\frac{d}{dx}\ln (I(x)-1),
\ee
and 
\be\label{am}
V^{[AM]}(x)=\hat{V}^{(-)}(\lambda=-1,x)=V^{(-)}(x)-2\frac{d^2}{dx^2}\ln (I(x)-1).
\ee
Thus as in the Pursey case, the supersymmetry between $V^{(+)}(x)$ and $V^{[P]}(x)$ is broken, thus the energy eigenvalues 
will be 
\be
E^{[AM]}_n=E^{(+)}_n\,.
\ee
Further, the scattering amplitudes in one dimensional cases are  
\be\label{ram}
r^{[AM]}(k)=r^{(-)}(k)
\ee
\be\label{tam}
t^{[AM]}(k)=-\bigg(\frac{W_{+}+ik'}{W_{+}-ik'}\bigg)t^{(-)}(k)
\ee
while in the three dimensional case  
\be\label{scattam}
s^{[AM]}(k')=\bigg(\frac{W_{+}+ik'}{W_{+}-ik'}\bigg)^2 s^{(-)}(k')\,.
\ee
In this case, the normalized eigenfunctions $\psi^{[AM]}_{n}(x)$ 
($n=0,1,2,...$) for the 
 AM potentual $V^{[AM]}(x)$ are obtained by putting $\lambda=-1$ in 
Eq. (\ref{extwf}), i.e. 
\ba\label{extwfam}
\psi^{[AM]}_{n}(x)&=&\hat{\psi}^{(-)}_{n+1}(\lambda=-1,x)\nonumber\\
&=&\psi^{(-)}_{n+1}(x)+\frac{1}{E^{(-)}_{n+1}}\bigg(\frac{I'(x)}{I(x)-1}\bigg)\bigg( \frac{d}{dx}+W(x)\bigg)\psi^{(-)}_{n+1}(x).
\ea
Summarizing, given any potential $V^{(-)}(x)$ with at least one bound state, one
can always construct three potentials $V^{(+)}, V^{[P]}(x), V^{[AM]}(x)$ which are
isospctral to each other but may or may not be strictly isospectral, i.e.
while their energy eigenvalues are the same, their reflection and transmission 
amplitudes in the one dimensional case or scattering amplitude in the three
dimensional case may or may not be the same. For example, for potentials with
purely discrete spectrum, obviously they are strictly isospectral.

\section{Rationally extended isospectral potentials}
In this section,  we consider three examples of RE exactly solvable shape 
invariant potentials (namely 
the extended radial oscillator, extended Scarf-I and extended generalized 
P\"oschl-Teller potentials) whose solutions 
are associated with the $X_m$ Laguerre (extended radial oscillator) and 
$X_m$ Jacobi (extended  Scarf-I 
and generalized P\"oschl-Teller potentials)EOPs \cite{os,midya,rkyd} and in all
three cases obtain explicitly 
the corresponding rationally extended one-parameter family of isospectral 
potentials. Since we are 
considering the RE potentials which are $m$-dependent (the corresponding 
eigenfunctions being $X_m$ Laguerre or $X_m$ Jacobi polynomials), 
hence we introduce a subscript $m$ with all the 
terms defined in the last section such as  $W(x)\Rightarrow W_{m}(x)$, $V^{(\pm)}(x)\Rightarrow V^{(\pm)}_m(x)$, $I(x)\Rightarrow I_m(x)$ 
etc. For all these extended potentials, as is well known, 
the total superpotential is the sum of the superpotential of the conventional 
(non-rational) term 
($W_{con}(x)$) and the superpotential due to the rational term ($W_{m,rat}(x)$)
, i.e.  
\be\label{ratsp}
W_{m}(x)=W_{con}(x)+W_{m,rat}(x),
\ee
Thus the rationally extended potential becomes 
\be\label{repot}
V^{(-)}_m(x) =W^2_{m}(x)-W'_{m}(x)=V^{(-)}_{con}(x)+V^{(-)}_{m,rat}(x),
\ee
Here $V^{(-)}_{con}(x)$ is the corresponding conventional potential and 
$V^{(-)}_{m,rat}(x)$ is the extra piece due to the rational term. 
Similarly the partner potential $V^{(+)}_m(x)$ with the associated usual and 
rational partner terms is given by
\be\label{prat}
V^{(+)}_m(x) =W^2_{m}(x)+W'_{m}(x)=V^{(+)}_{con}(x)+V^{(+)}_{m,rat}(x).
\ee
For $m=0$, the rational superpotential  term $W_{m,rat}(x)=0$ and hence $V^{(-)}_{m,rat}(x)=0$. Thus the extended potential
 $V^{(-)}_m(x)$ reduces to the conventional potential $V^{(-)}_{con}(x)$. 

\subsection{Rationally extended radial oscillator potential}
In this case the terms associated with the superpotenial (\ref{ratsp}) are given by \cite{os,rkyd}
\be
W_{con}(r)=\frac{\omega r}{2}-\frac{(\ell+1)}{r}
\ee 
and 
\be
W_{m,rat}(r)=\omega r\bigg[\frac{L^{(\ell+\frac{1}{2})}_{m-1}(-\frac{\omega r^2}{2})}{L^{(\ell-\frac{1}{2})}_{m}(-\frac{\omega r^2}{2})}-\frac{L^{(\ell+\frac{3}{2})}_{m-1}(-\frac{\omega r^2}{2})}{L^{(\ell+\frac{1}{2})}_{m}(-\frac{\omega r^2}{2})}\bigg].
\ee
Thus the terms associated with the RE potential (\ref{repot}) become 
\be
V^{(-)}_{con}(r)=\frac{1}{4}\omega^2 r^2+\frac{\ell(\ell+1)}{r^2}-\omega(\ell+\frac{3}{2})
\ee
and 
\ba
V^{(-)}_{m,rat}(r)&=&-\omega^2r^2\frac{L^{(\ell+\frac{3}{2})}_{m-2}(-\frac{\omega r^2}{2})}{L_{m}^{(\ell-\frac{1}{2})}(-\frac{\omega r^2}{2})}+\omega(\omega r^2+2\ell-1)\frac{L^{(\ell+\frac{1}{2})}_{m-1}(-\frac{\omega r^2}{2})}{L_{m}^{(\ell-\frac{1}{2})}(-\frac{\omega r^2}{2})}\nonumber\\
&+&2\omega^2 r^2\bigg(\frac{L^{(\ell+\frac{1}{2})}_{m-1}(-\frac{\omega r^2}{2})}{L_{m}^{(\ell-\frac{1}{2})}(-\frac{\omega r^2}{2})}\bigg)^2 - 2m\omega,\quad 0<r<\infty.
\ea
respectively. Similarly, the terms associated with the partner potential (\ref{prat}) are 
\be
V^{(+)}_{con}(r)=\frac{1}{4}\omega^2 r^2+\frac{(\ell+1)(\ell+2)}{r^2}-\omega(\ell+\frac{3}{2})
\ee
and 
\ba
V^{(+)}_{m,rat}(r)&=&-\omega^2r^2\frac{L^{(\ell+\frac{5}{2})}_{m-2}(-\frac{\omega r^2}{2})}{L_{m}^{(\ell+\frac{1}{2})}(-\frac{\omega r^2}{2})}+\omega(\omega r^2+2\ell+1)\frac{L^{(\ell+\frac{3}{2})}_{m-1}(-\frac{\omega r^2}{2})}{L_{m}^{(\ell+\frac{3}{2})}(-\frac{\omega r^2}{2})}\nonumber\\
&+&2\omega^2 r^2\bigg(\frac{L^{(\ell+\frac{3}{2})}_{m-1}(-\frac{\omega r^2}{2})}{L_{m}^{(\ell+\frac{1}{2})}(-\frac{\omega r^2}{2})}\bigg)^2 - 2m\omega.
\ea
The ground state eigenfunction corresponding to the extended potential 
$V^{(-)}_m(r)$ in terms of $X_m$ Laguerre polynomials
$\hat{L}_{m}^{(\ell+\frac{1}{2})}(\frac{\omega r^2}{2})$ can be obtained using 
Eq. (\ref{sup}) and one gets
\be\label{3.2}
\psi^{(-)}_{0,m}(r) = N^{\ell}_{m}\frac{r^{\ell+1}\exp\big(-\frac{\omega r^2}{4}\big)}{L^{(\ell-\frac{1}{2})}_m(-\frac{\omega r^2}{2})}\hat{L}_{m}^{(\ell+\frac{1}{2})}(\frac{\omega r^2}{2}), \quad m=1,2,...,
\ee
where $N^{\ell}_{m}$ is the normalization constant which is given by 
\be\label{3.5}
N_{m}=\bigg[\frac{\omega^{(\ell+\frac{3}{2})}}{2^{(\ell+\frac{1}{2})}(\ell+2m-\frac{1}{2})\Gamma(\ell+m-\frac{1}{2})}\bigg]^{1/2}.
\ee 
The excited state eigenfunctions ($n=0,1,2..$) of $V^{(-)}_m(r)$ 
are also explicitly 
known \cite{os} and given by 
\be\label{wfrm}
\psi^{(-)}_{n,m}(r) = N^{\ell}_{n,m}\frac{r^{\ell+1}\exp\big(-\frac{\omega r^2}{4}\big)}{L^{(\ell-\frac{1}{2})}_m(-\frac{\omega r^2}{2})}\hat{L}_{n+m}^{(\ell+\frac{1}{2})}(\frac{\omega r^2}{2}), \quad m=1,2,...,
\ee
and the energy eigenvalues are ($n = 0, 1, 2,..$)
\be\label{evr}
E^{(-)}_{n+1}= E^{+}_{n} = 2(n+1) \omega\,,~~E^{-}_0 = 0\,, 
\ee
with
\be
N^{\ell}_{n,m}=\bigg[\frac{n!\omega^{(\ell+\frac{3}{2})}}{2^{(\ell+\frac{1}{2})}(\ell+n+2m-\frac{1}{2})\Gamma(\ell+n+m-\frac{1}{2})}\bigg]^{1/2}.
\ee 

Hence the eigenfunctions of the partner potentials are 
 \be\label{wfprm}
\psi^{(+)}_{n,m}(r) = N^{\ell+1}_{n,m}\frac{r^{\ell+2}\exp\big(-\frac{\omega r^2}{4}\big)}{L^{(\ell+\frac{1}{2})}_m(-\frac{\omega r^2}{2})}\hat{L}_{n+m}^{(\ell+\frac{3}{2})}(\frac{\omega r^2}{2}).
\ee
In terms of the classical Laguerre polynomials the $X_m$ Laguerre polynomials 
can be written as
\be
\hat{L}^{(\alpha)}_{n+m}(z)=L^{(\alpha)}_m(-z)L^{(\alpha-1)}_{n}(z)+L^{(\alpha-1)}_m(-z)L^{(\alpha)}_{n-1}(z); \quad n\geq m.
\ee

Using the ground state eigenfunction as give by Eqs. (\ref{3.2}) and 
(\ref{3.5}) and the results discussed in 
the last section, it is straight forward to calculate $I_{m}(r)$ as defined 
by Eq. (\ref{intim}) and obtain 
one continuous parameter family of RE strictly isospectral potentials 
$\hat{V}^{(-)}_{m}(\lambda,r)$ 
and also the corresponding eigenfunctions $\psi^{(-)}_{n,m}(\lambda, r)$ 
in terms of 
$X_m$ Laguerre polynomials. 
In the particular cases of $\lambda=0$ and  $\lambda = -1$ one then obtains
the extended pursey $V^{[P]}_m(r)$  and the 
extended Abraham-Moses $V^{[AM]}_m(r)$ potentials respectively. 

We now illustrate this general approach by explicitly discussing the $X_1$ 
case ($m=1$). In 
this case the extended potential (\ref{repot}) and the corresponding 
 eigenfunctions (\ref{wfrm}) are 
\be\label{3.4}
V^{(-)}_{m=1}(r)=\frac{1}{4}\omega r^2+\frac{\ell(\ell+1)}{r^2}+4\omega \bigg(\frac{1}{(\omega r^2+2\ell+1)}-\frac{2(2\ell+1)}{(\omega r^2+2\ell+1)^2}\bigg)-\omega(\ell+\frac{3}{2})
\ee
and 
\be\label{3.3}
\psi^{(-)}_{n,m = 1}(r) = N^{\ell}_{n,1}\frac{r^{\ell+1}\exp\big(-\frac{\omega r^2}{4}\big)}{L^{(\ell-\frac{1}{2})}_1(-\frac{\omega r^2}{2})}\hat{L}_{n+1}^{(\ell+\frac{1}{2})}(\frac{\omega r^2}{2})
\ee
respectively. The energy eigenvalues are the same as that of the conventional 
case and are given by Eq. (\ref{evr}).
Similarly from Eq. (\ref{prat}), the corresponding partner potential is
\be
V^{(+)}_{1}(r)=\frac{1}{4}\omega r^2+\frac{(\ell+1)(\ell+2)}{r^2}+4\omega \bigg(\frac{1}{(\omega r^2+2\ell+3)}-\frac{2(2\ell+3)}{(\omega r^2+2\ell+3)^2}\bigg)-\omega(\ell+\frac{3}{2}),
\ee
and the corresponding eigenfunctions  can be easily obtained from
those of $\psi^{(+)}_{n, m=1}$ as given by Eq. (\ref{wfprm}). 

As an illustration, we have calculated $I_{m=1}(r)$ in the case of 
$\omega=2$ and $\ell=1$ and is given by  
\be\label{intrad}
I_1(r)=-\frac{2 \exp (-r^2)r(15 + 4r^2(5+r^2))}{5\sqrt{\pi} (3 + 2r^2)} 
+ \erf (r),
\ee
where $\erf(r)$ is an error function. Using this expression of $I_1(r)$,
one immediately obtains one continuous parameter family $\hat{V}^{(-)}(\lambda, r)$ 
of potentials which are strictly isospectral to the potential $V^{(-)}(r)$ as
given by Eq. (\ref{3.4}) in case $\omega = 2$ and $l = 1$ and provided if
$\lambda > 0$ or $\lambda < -1$. The corresponding eigenfunctions are then
easily obtained from Eqs. (\ref{gswf}) and (\ref{extwf}). 

One can also obtain the corresponding RE Pursey (\ref{pur}) and the RE 
AM (\ref{am}) potentials by using the formula
\be\label{prx1}
V^{[P]}_1(r)=V^{(-)}_{1}(r)-2\frac{d^2}{dr^2}\ln (I_1(r))
\ee
and
\be\label{amrx1}
V^{[AM]}_1(r)=V^{(-)}_{1}(r)-2\frac{d^2}{dr^2}\ln (I_1(r)-1)
\ee
 respectively, where $I_1(r)$ is as given by Eq. (\ref{intrad}). 
The corresponding eigenfunctions can be easily obtained by setting $m=1$ in Eqs. (\ref{extwfp}) and (\ref{extwfam}) respectively. 

Using the expression of $I_1(r)$ as given by Eq. (\ref{intrad}),
 we have plotted some of the RE isospectral potentials  and the corresponding
ground state eigenfunctions in Figs $1(a)$ to $1(d)$.
In particular, in Fig. $1(a)$, we have considered the variation of $\lambda$ 
from $+\infty$ to zero. 
In this case as $\lambda$ decreases from $+\infty$, the potential starts 
developing a minima and the corresponding position starts shifting towards 
$r=0$. The limiting case of $\lambda=0$ gives the corresponding  
rationally extended Pursey potential $V^{[P]}_{1}(r)$, whose solutions are 
associated with $X_1$ Laguerre EOPs.
In Fig. $1(b)$, we have plotted the potentials for various values of $\lambda$ 
 from $-\infty$ to $-1$.  
As $\lambda$ increases from $-\infty$
and approaches $-1$,  the resulting attractive potential well shifts 
towards 
$r=\infty$ and in the limiting case of $\lambda=-1$, we get the rationally 
extended Abraham-Moses potential $V^{[AM]}_{1}(r)$.
In Fig. $1(c)$ we have plotted  the three strictly isospectral potentials 
namely the potential $V^{(+)}_{m=1}(r)$, the RE Pursey potential 
$V^{[P]}_{1}(r)$ and the 
RE Abraham-Moses potential $V^{[AM]}_{1}(r)$.
The corresponding ground state eigenfunction for all the strictly isospectral 
RE potentials with positive $\lambda$
drawn in Fig. $1(a)$ are plotted in Fig. $1(d)$. 

\includegraphics[scale=1.6]{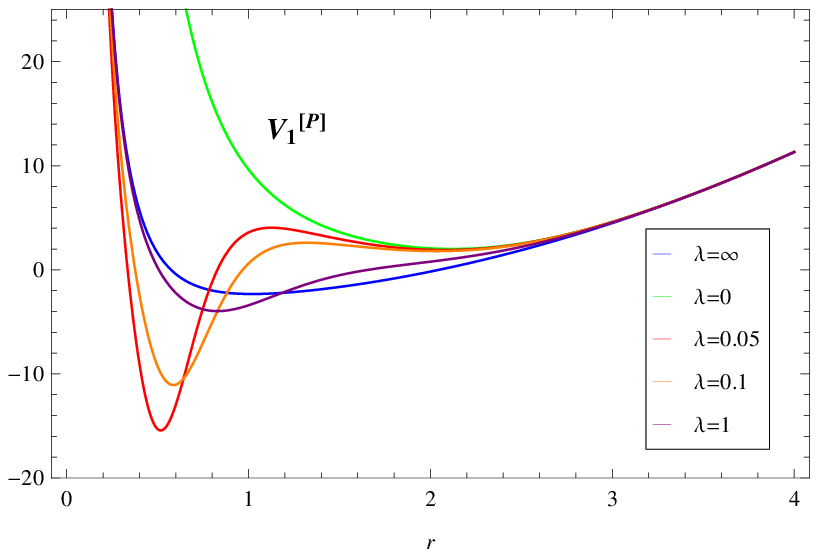}\\ 
{\bf Fig.1}: {(a) {\it Rationally extended potentials $\hat{V}_1(\lambda,r)$ strictly isospectral to the extended 
$3D$ oscillator potential $V^{(-)}_1(r)$ for positive $\lambda=0,0.05,0.1,1$ and $\infty$. The extended Pursey potential is shown for $\lambda=0$.}\\\
\includegraphics[scale=1.65]{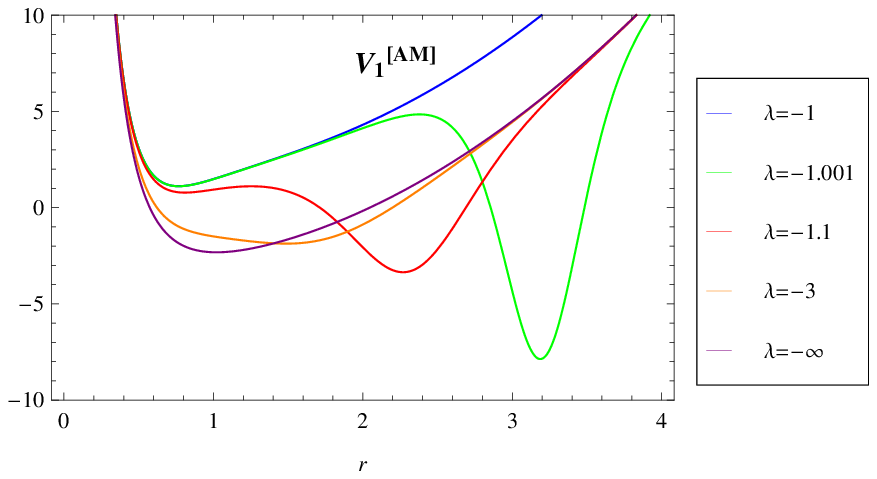} \\
{\bf Fig.1}: {(b) {\it Rationally extended potentials $\hat{V}_1(\lambda,r)$ 
for negative $\lambda=-\infty,-3,-1.1,-1.001$ and $-1$. The extended AM potential is shown for $\lambda=-1$.}\\\ 
 \includegraphics[scale=1.6]{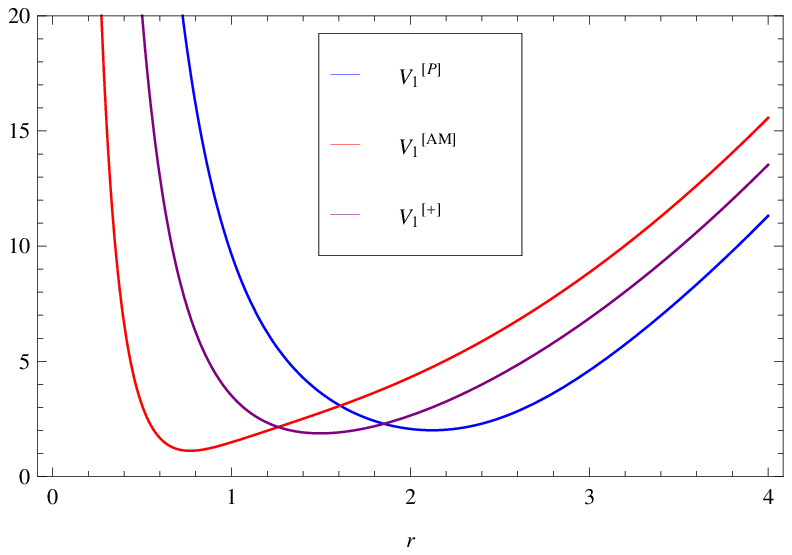}\\
{\bf Fig.1}: {(c) {\it The extended Pursey potential $V^{[P]}_1(r)$, extended AM potential $V^{[AM]}_1(r)$ and the 
extended partner potentials $V^{(+)}_1(r)$.}\\\
 \includegraphics[scale=1.6]{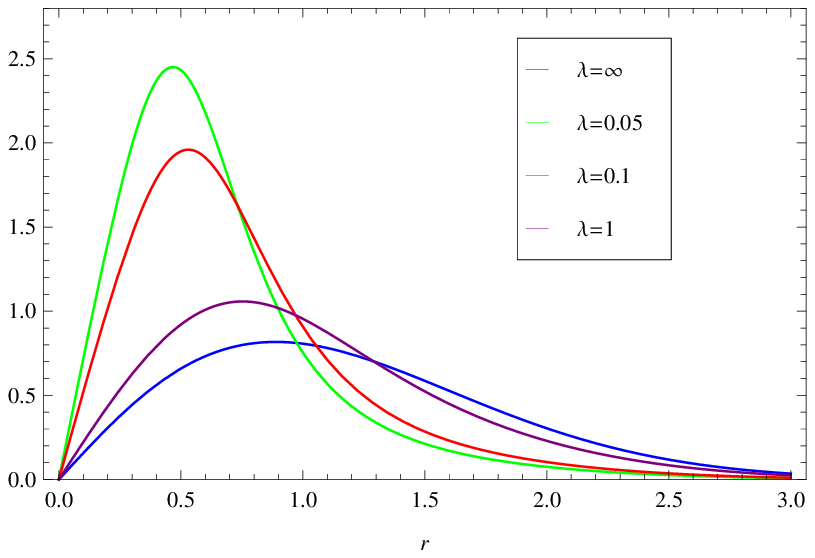}\\
{\bf Fig.1}: {(d) {\it Normalized ground-state wavefunctions $\frac{\hat{\psi}_{0,1}(\lambda,r )}{r}$ for some potentials 
 (with positive $\lambda$)shown in Fig. $1(a)$. }\\
   

\subsection{Rationally Extended Scarf-I potential} 

In this case the terms in the superpotential (\ref{ratsp}) are defined as \cite{midya}
\be\label{resc}
W_{con}(x) =A\tan{x} -B\sec{x}
\ee
and 
\be
W_{m,rat}(x) =-\frac{(\beta-\alpha+m-1)}{2}\cos x\bigg[\frac{P^{(-\alpha-1,\beta+1)}_{m-1}(z)}{P^{(-\alpha-2,\beta)}_{m}(z)}-\frac{P^{(-\alpha,\beta)}_{m-1}(z)}{P^{(-\alpha-1,\beta-1)}_{m}(z)}\bigg],
\ee
where $\alpha=A-B-\frac{1}{2}$, \qquad $\beta=A+B-\frac{1}{2}$ and $P^{(\alpha,\beta)}_{m}(z)$ (here $z=\sin x $) is a classical Jacobi polynomial. 
Using $W_{con}(x)$ and $W_{m,rat}(x)$ in (\ref{repot}), we get the extended 
Scarf-I potential $V^{(-)}_{m}(x)$  with the terms
\be\label{scarf}
V^{(-)}_{con}(x)= [(A-1)A+B^{2}]\sec^{2}{x}-B(2A-1)\sec{x}\tan{x}-A^{2}
\ee
and 
\ba
V_{m,rat}^{(-)}(x)&=&(2B-m-1)[2A-1+(-2B+1)\sin x]\bigg(\frac{P^{(-\alpha,\beta)}_{m-1}(\sin x)}{P^{(-\alpha-1,\beta-1)}_{m}(\sin x)}\bigg)\nonumber\\
&+&\frac{(-2B-m+1)^2}{2}\cos^2x \bigg(\frac{P^{(-\alpha,\beta)}_{m-1}(\sin x)}{P^{(-\alpha-1,\beta-1)}_{m}(\sin x)}\bigg)^2\nonumber\\
&-&2m(-2B-m-1); \quad -\pi/2<x<\pi/2,\quad 0<B<A-1,
\ea
while the terms associated with the partner potential $V^{(+)}_{m}(x)$ are 
\be\label{scarfp}
V^{(+)}_{con}(x)= [(A+1)A+B^{2}]\sec^{2}{x}-B(2A+1)\sec{x}\tan{x}-A^{2}
\ee
and 
\ba
V_{m,rat}^{(+)}(x)&=&(2B-m-1)[2A+1+(-2B+1)\sin x]\bigg(\frac{P^{(-\alpha-1,\beta+1)}_{m-1}(\sin x)}{P^{(-\alpha-2,\beta)}_{m}(\sin x)}\bigg)\nonumber\\
&+&\frac{(-2B-m+1)^2}{2}\cos^2x \bigg(\frac{P^{(-\alpha-1,\beta+1)}_{m-1}(\sin x)}{P^{(-\alpha-2,\beta)}_{m}(\sin x)}\bigg)^2\nonumber\\
&-&2m(-2B-m-1).
\ea
Using Eq. (\ref{sup}) the normalized ground state eigenfunction corresponding to the rationally extended Scarf-I potential $V^{(-)}_{m}(x)$ is 
given by  
\be\label{wfsc}
\psi^{(-)}_{0,m}(x)=N^{(\alpha,\beta)}_{m}\frac{(1-\sin x)^{\frac{(A-B)}{2}}(1+\sin x)^{\frac{(A+B)}{2}}}{P^{(-\alpha-1,\beta-1)}_{m}(\sin x)}\hat{P}^{(\alpha,\beta)}_{m}(\sin{x}),\quad m=1,2...
\ee 
where the normalization constant 
\be\
N^{(\alpha,\beta)}_{m}=\bigg(\frac{(\alpha+1)\Gamma(\alpha+\beta+2)}{2^{\alpha+\beta+1}(\alpha-m+1)(m+\beta)\Gamma(\alpha+1)\Gamma(\beta)}\bigg)^{1/2}.
\ee
The excited state eigenfunctions are also known \cite{midya} and are given by 
\be\label{wfscm}
\psi^{(-)}_{n,m}(x)=N^{(\alpha,\beta)}_{n,m}\frac{(1-\sin x)^{\frac{(A-B)}{2}}(1+\sin x)^{\frac{(A+B)}{2}}}{P^{(-\alpha-1,\beta-1)}_{m}(\sin x)}\hat{P}^{(\alpha,\beta)}_{n+m}(\sin{x})
\ee 
The energy eigenvalues are ($n = 0, 1, 2,...$)  
\be\label{ensc}
E^{(-)}_{n+1}= E^{+}_{n} = (A+n+1)^2 - A^2\,, ~~ E^{-}_0 = 0\,,
\ee 
while the normalization constant is
\be
N^{(\alpha,\beta )}_{n,m}=\bigg[\frac{n!(n+\alpha+1)^2(\alpha+\beta+2n+1)\Gamma(n+\alpha+\beta+1)}{2^{\alpha+\beta+1}(n+\alpha-m+1)(n+m+\beta)\Gamma(n+\alpha+2)\Gamma(n+\beta)}\bigg]^{\frac{1}{2}}.
\ee
The wavefunctions associated with the partner potentials can be obtained from 
Eq. (\ref{pwf}) and gets 
\be\label{wfscpm}
\psi^{(+)}_{n,m}(x)=N^{(\alpha+1,\beta+1)}_{n,m}\frac{(1-\sin x)^{\frac{(A-B+1)}{2}}(1+\sin x)^{\frac{(A+B+1)}{2}}}{P^{(-\alpha-2,\beta)}_{m}(\sin x)}\hat{P}^{(\alpha+1,\beta+1)}_{n+m}(\sin{x}),
\ee 
where the exceptional Jacobi polynomial $\hat{P}^{(\alpha,\beta)}_{n+m}(z)$ in terms of the classical Jacobi polynomials is written as
\ba
\hat{P}^{(\alpha,\beta)}_{n+m}(z)&=&(-1)^m\bigg[\bigg(\frac{1+\alpha+\beta+n}{1+\alpha+n}\bigg)\bigg(\frac{z-1}{2}\bigg)P^{(-\alpha-1,\beta-1)}_{m}(z)P^{(\alpha+2,\beta)}_{n-1}(z)\nonumber\\
&+&\bigg(\frac{1+\alpha-m}{1+\alpha+n}\bigg)P^{(-2-\alpha,\beta)}_{m}(z)P^{(\alpha+1,\beta-1)}_{n}(z)\bigg].
\ea

Using the normalized ground state eigenfunction as given by Eqs. (\ref{wfsc}), 
it is now straight forward to calculate the indefinite 
integral
$I_{m}(x)$ by using the definition (\ref{intim}). 
Using this expression of $I_{m}(x)$ it is then 
straight forward to obtain one continuous parameter family of RE potentials 
$\hat{V}^{-}_m(\lambda, x)$ which are strictly isospectral to the potential 
$V^{-}_m(x)$.  
One can also obtain the corresponding RE Pursey (\ref{pur}) and the RE 
AM (\ref{am}) potentials by using the Eqs. (\ref{prx1}) and (\ref{amrx1}). 
 respectively.  
The corresponding eigenfunctions can be easily obtained by using 
Eqs. (\ref{extwfp}) and (\ref{extwfam}) respectively.

As an illustration, we now consider the $X_1$ case ($m=1$)  in detail. In this
case the potentials and the wavefunctions are written as 
\ba\label{3.7}
V^{(-)}_1(x) &= &[(A-1)A+B^2]\sec^{2}{x}-B(2A-1)\sec{x}\tan{x}\nonumber\\
&+&2\bigg(\frac{(2A-1)}{(2A-1-2B\sin x)}-\frac{[(2A-1)^2-B^2]}{(2A-1-2B\sin x)^2}\bigg)-A^2
\ea
and 
\be\label{wfscx1}
\psi^{(-)}_{n,1}(x)=N^{(\alpha,\beta)}_{n,1}\frac{(1-\sin x)^{\frac{(A-B)}{2}}(1+\sin x)^{\frac{(A+B)}{2}}}{P^{(-\alpha-1,\beta-1)}_{1}(\sin x)}\hat{P}^{(\alpha,\beta)}_{n+1}(\sin{x})
\ee
respectively. The energy eigenvalues are same as that of the $X_m$ case (\ref{ensc}). 
In this case, the partner potential (\ref{prat}) is given by
\ba
V^{(+)}_1(x) &=& [(A+1)A+B^2]\sec^{2}{x}-B(2A+1)\sec{x}\tan{x}\nonumber\\
&+&2\bigg(\frac{(2A+1)}{(2A+1-2B\sin x)}-\frac{[(2A+1)^2-B^2]}{(2A+1-2B\sin x)^2}\bigg)-A^2.
\ea
 The associated wavefunction $\psi^{(+)}_{n,1}(x)$ can also be easily obtain from Eq. (\ref{wfscpm}).
Using these results, the RE Pursey (\ref{pur}) and the extended 
AM (\ref{am}) potentials with their corresponding wavefunctions (\ref{extwfp}) and (\ref{extwfam}) 
in terms of  $X_1$ Jacobi EOPs can be find by 
setting $m=1$. 

 
 

As an illustration, we have calculated $I_{m=1}(x)$ in the case of 
$A = 3, B = 1$ and is given by  
\ba\label{intscarf}
I_1(x)=\frac{1}{2}&+&\frac{1}{180\pi(-5+2\sin x)}[-900 x +675 \cos x +176 \cos 3x+44 \cos 5x \nonumber\\
&+&\cos 7x+360x\sin x-575\sin 2x-55\sin 4x+5\sin 6x].
\ea
Using this expression of $I_1(x)$,
one immediately obtains one continuous parameter family $\hat{V}^{(-)}_1(\lambda, x)$ 
of potentials which are strictly isospectral to the potential $V^{(-)}_1(x)$ as
given by Eq. (\ref{3.7}) in case $A = 3, B = 1$ and provided if
$\lambda > 0$ or $\lambda < -1$. The corresponding eigenfunctions are then
easily obtained too. 
One can also obtain the corresponding RE Pursey (\ref{pur}) and the RE 
AM (\ref{am}) potentials by using the Eqs. (\ref{prx1}) and (\ref{amrx1}).
  
Using the above $I_1(x)$, we have plotted some of these extended isospectral 
potentials and normalized ground state wavefunctions in Figs. $2$(a)-$2$(d). 
 In this case as $\lambda$ varies from $\infty$ to $0$ (Fig. $2$(a)), the minima starts shifting towards 
$x=0$ and as $\lambda$ reduces to zero, we get the rationally extended Pursey potential. 
Similarly, in Fig. $2$(b) we have plotted the potentials 
for negative $\lambda$ and as $\lambda$ approaches $-1$, the minima shifts 
towards $x=\infty$. Ultimately
as $\lambda=-1$, we get the rationally extended AM potential. Plots of three 
strictly isospectral potentials namely the RE partner potential 
$V^{(+)}_1(x)$, the RE Pursey potential $V^{[P]}_1$ and the RE AM potential 
$V^{[AM]}_1$ are 
shown in Fig. $2$ (c). The normalized ground state eigenfunctions  
$\hat{\psi}^{(-)}_{0,1}(x)$ of all the potentials with positive $\lambda$ 
as shown in Fig. $2$ (a) are also plotted in Fig. $2$ (d). 

\includegraphics[scale=1.6]{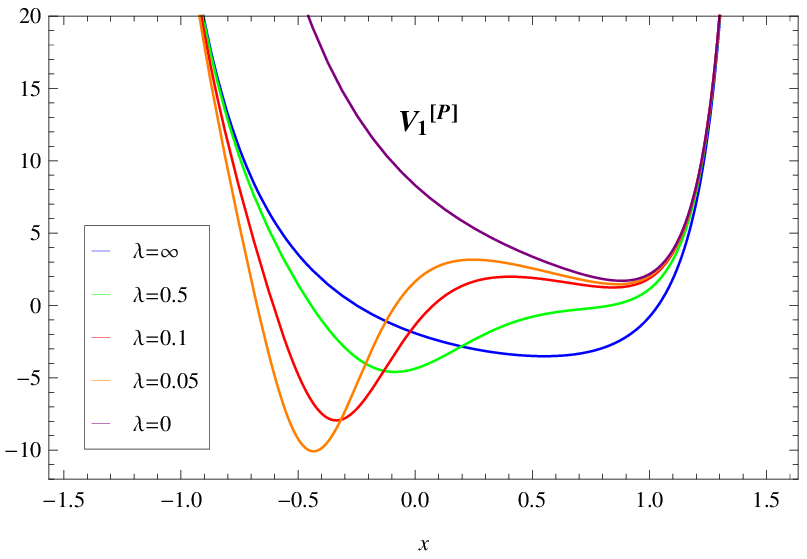}\\ 
 {\bf Fig.2}: {(a) {\it Rationally extended potentials $\hat{V}_1(\lambda,x)$ strictly isospectral to the extended 
Scarf-I potential $V^{(-)}_1(x)$ for positive $\lambda=0, 0.05,0.1,0.5$ and $\infty$.}\\
\includegraphics[scale=1.7]{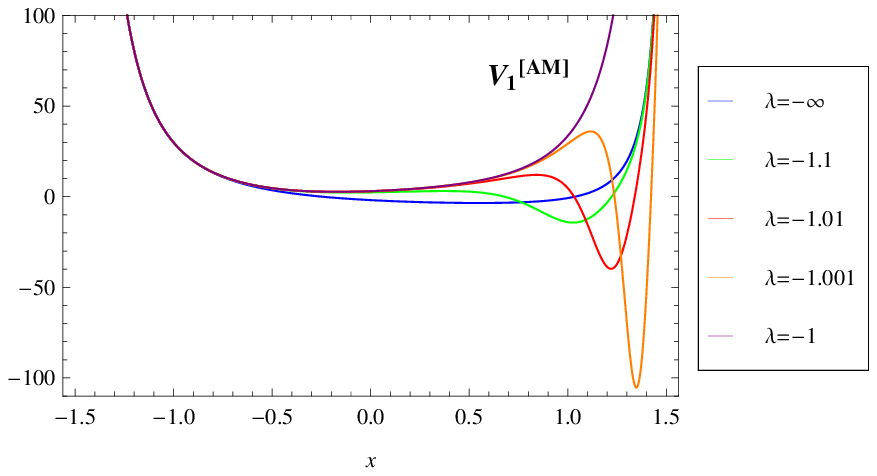}\\
{\bf Fig.2}: {(b) {\it Rationally extended potentials $\hat{V}_1(\lambda,x)$ 
for negative $\lambda=-\infty, -1.1,-1.01, -1.001$ and $-1$.} \\
  \includegraphics[scale=1.6]{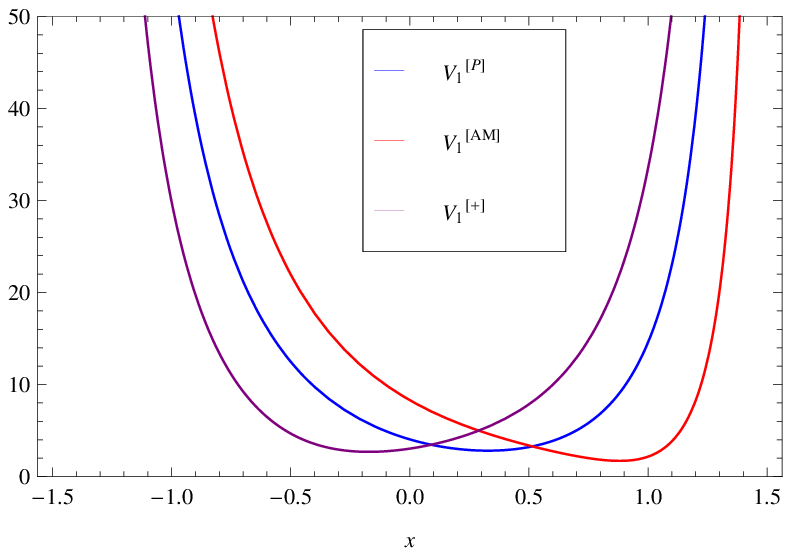}\\
  {\bf Fig.2}: {(c) {\it The extended Pursey potential $V^{[P]}_1(x)$, 
extended AM potential $V^{[AM]}_1(x)$ and the 
extended partner potentials $V^{(+)}_1(x)$.}\\
 \includegraphics[scale=1.6]{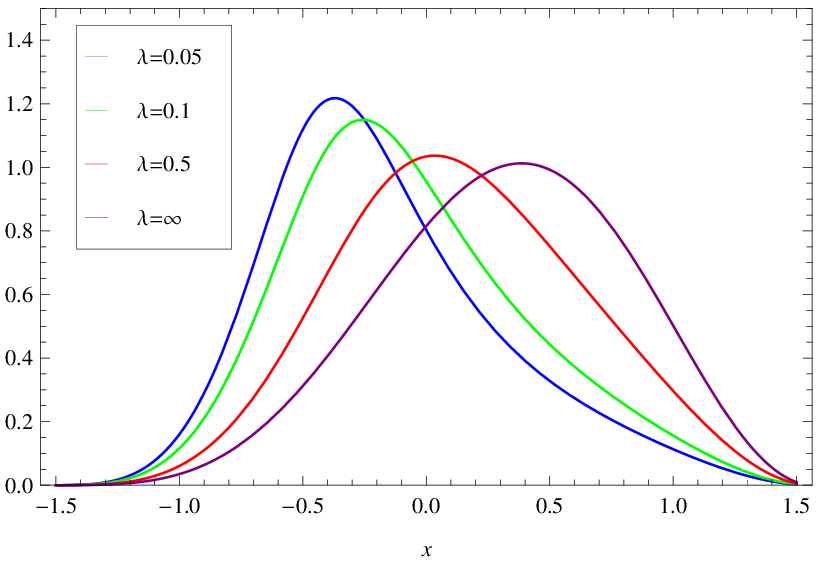}\\
  {\bf Fig.2}:{(d) {\it Normalized ground-state eigenfunctions $\hat{\psi}_{0,1}(\lambda,x )$ for some of the potentials 
with positive $\lambda$ as shown in Fig. $2(a)$. }\\

 
 

\subsection{Rationally Extended generalized Poschl-Teller (GPT) potential}
As a last example, we consider the RE GPT potential which has both the bound as well as the scattering 
state solutions. The terms corresponding to the superpotential (\ref{ratsp}) associated with this potential are given as  
\be\label{wugpt}
W_{con}(x) =A\coth{x} -B\cosech{x}
\ee
and 
\be\label{wrgpt}
W_{m,rat}(x) =-\frac{(\beta-\alpha+m-1)}{2}\sinh x\bigg[\frac{P^{(-\alpha-1,\beta+1)}_{m-1}(z)}{P^{(-\alpha-2,\beta)}_{m}(z)}-\frac{P^{(-\alpha,\beta)}_{m-1}(z)}{P^{(-\alpha-1,\beta-1)}_{m}(z)}\bigg],
\ee
where $\alpha=B-A-\frac{1}{2}$, \qquad $\beta=-B-A-\frac{1}{2}$ and $z=\cosh x $. 
Using (\ref{wugpt})and (\ref{wrgpt}) in (\ref{repot}), we get the extended GPT potential $V^{(-)}_{m}(x)$  with the terms
\be\label{gpt}
V^{(-)}_{con}(x)= [(A+1)A+B^{2}]\cosech^{2} x-B(2A+1)\cosech x\coth x+A^{2}
\ee
and 
\ba
V_{m,rat}^{(-)}(x)&=&-(2B-m+1)[2A+1-(2B+1)\cosh x]\bigg(\frac{P^{(-\alpha,\beta)}_{m-1}(\cosh x)}{P^{(-\alpha-1,\beta-1)}_{m}(\cosh x)}\bigg)\nonumber\\
&+&\frac{(2B-m+1)^2}{2}\sinh^2x \bigg(\frac{P^{(-\alpha,\beta)}_{m-1}(\cosh x)}{P^{(-\alpha-1,\beta-1)}_{m}(\cosh x)}\bigg)^2\nonumber\\
&+&2m(-2B-m-1); \quad 0\leq x\leq \infty, \quad B>A+1>1,
\ea
while the terms associated with the partner potential $V^{(+)}_{m}(x)$ are 
\be\label{gptp}
V^{(+)}_{con}(x)= [A(A-1)+B^{2}]\cosech^{2} x-B(2A-1)\cosech x\coth x+A^{2}
\ee
and 
\ba
V_{m,rat}^{(+)}(x)&=&-(2B-m+1)[2A+1-(2B+1)\cosh x]\bigg(\frac{P^{(-\alpha-1,\beta+1)}_{m-1}(\cosh x)}{P^{(-\alpha-2,\beta)}_{m}(\cosh x)}\bigg)\nonumber\\
&+&\frac{(2B-m+1)^2}{2}\sinh^2x \bigg(\frac{P^{(-\alpha-1,\beta+1)}_{m-1}(\cosh x)}{P^{(-\alpha-2,\beta)}_{m}(\cosh x)}\bigg)^2\nonumber\\
&+&2m(-2B-m-1).
\ea
The normalized ground state eigenfunction corresponding to the rationally 
extended GPT potential $V^{(-)}_{m}(x)$ is easily calculated  and is given by
\be\label{wfgpt}
\psi^{(-)}_{0,m}(x)=N^{(\alpha,\beta)}_{m}\frac{(\cosh x-1)^{\frac{(B-A)}{2}}(\cosh x+1)^{-\frac{(B+A)}{2}}}{P^{(-\alpha-1,\beta-1)}_{m}(\cosh x)}\hat{P}^{(\alpha,\beta)}_{m}(\cosh x),\quad m=1,2...
\ee 
where the normalization constant
\be
N^{(\alpha,\beta)}_{m}=\bigg[\frac{(-\alpha-\beta-1)(\alpha+1)\Gamma(-\beta+1)}{2^{\alpha+\beta+1}(-\beta-m)(\alpha-m+1)\Gamma(\alpha+1)\Gamma(-\alpha-\beta)}\bigg]^{1/2}.
\ee
The excited state solutions  are also known \cite{scatt}
and given as 
\be\label{wfgptm}
\psi^{(-)}_{n,m}(x)=N^{(\alpha,\beta)}_{n,m}\frac{(\cosh x-1)^{\frac{(B-A)}{2}}(\cosh x+1)^{-\frac{(B+A)}{2}}}{P^{(-\alpha-1,\beta-1)}_{m}(\cosh x)}\hat{P}^{(\alpha,\beta)}_{n+m}(\cosh x),\quad m=1,2...
\ee 
The energy eigenvalues are given by ($n = 0, 1, 2$) 
\be\label{engpt}
E^{(-)}_{n+1}= E^{(+)}_{n} = A^2-(A-n-1)^2\,,~~E^{(-)}_{0} = 0\,, 
\ee
where the normalization constant
\be\label{gptnor}
N^{(\alpha,\beta)}_{n,m}=\bigg[\frac{n!(-\alpha-\beta-2n-1)(n+\alpha-m+1)(\alpha+n+1)\Gamma(-\beta-n+1)}{2^{\alpha+\beta+1}(-\beta-n-m)(\alpha-m+1)^2\Gamma(\alpha+n+1)\Gamma(-\alpha-\beta-n)}\bigg]^{1/2}.
\ee
The wavefunctions corresponding to the partner potential can be obtain using Eq. (\ref{pwf}) i.e,
\be\label{wfgptpm}
\psi^{(+)}_{n,m}(x)=N^{(\alpha+1,\beta+1)}_{n,m}\frac{(\cosh x-1)^{\frac{(B-A+1)}{2}}(\cosh x+1)^{-\frac{(B+A-1)}{2}}}{P^{(-\alpha-2,\beta)}_{m}(\cosh x)}\hat{P}^{(\alpha+1,\beta+1)}_{n+m}(\cosh x),\quad m=1,2...
\ee 
The scattering amplitudes ($s$-wave) related to the potential 
$V^{(-)}_{m}(x)$ are well known \cite{scatt} and are given by 
\ba\label{smgpt}
s^{(-)}_{m}(k')&=&2^{-4ik'}\frac{\Gamma(2ik')\Gamma(-A-ik')\Gamma(B-ik'+1/2)}{\Gamma(-2ik')\Gamma(-A+ik')\Gamma(B+ik'+1/2)}\nonumber\\
&\times&\Bigg[\frac{[B^2-(ik'-1/2)^2]+(B-ik'+1/2)(1-m)}{[B^2-(ik'+1/2)^2]+(B+ik'+1/2)(1-m)}\Bigg], \quad m\geq 1.
\ea

Just as in the last two examples, it is now straight forward to calculate 
$I_m(x)$ and hence one continuous parameter family of potentials 
$\hat{V}^{(-)}_m(\lambda, x)$ which are strictly isospectral to the RE GPT potential 
$V^{(-)}_m(x)$ in case $\lambda > 0$ or $\lambda < -1$. 
Further, The RE Pursey  and the RE AM
potentials  for this case can also be obtained in a straightforward way
in case $\lambda = 0$ or $\lambda = -1$ respectively. . 

From Eqs. (\ref{wugpt}) and (\ref{wrgpt}) as $x\rightarrow +\infty$, we get the asymptotic form of the superpotentials
\be
W_{con}(x\rightarrow +\infty)= A, \qquad W_{m,rat}(x\rightarrow +\infty)= 0,
\ee
hence the superpotential $W_m(x)$ becomes
\be
W_m(x\rightarrow +\infty) =A.
\ee
 Using Eqs. (\ref{scatt}), (\ref{scattp}) and (\ref{scattam}), the scattering 
amplitudes of the partner potential, 
the extended Pursey and the extended AM potentials are related to the
scattering amplitude of the $V^{(-)}_m(x)$ by 
\ba\label{3.10}
s^{(+)}_m(k')&=&\bigg(\frac{A+ik'}{A-ik'}\bigg)s^{(-)}_m(k')\nonumber\\
s^{[P]}_m(k')&=&s^{(-)}_m(k')\nonumber
\ea 
and
\be
s^{[AM]}_m(k')=\bigg(\frac{A+ik'}{A-ik'}\bigg)^2 s^{(-)}_m(k')
\ee
respectively. Thus unlike the two examples discussed above where the spectrum
was purely discrete and hence $V^{(+)}_m, V^{[P]}_m$ and $V^{[AM]}_m$ were strictly
isospectral, for RE GPT case, the three are isospectral but not 
strictly isospectral.  

We now illustrate the above results by discussing in detail the $X_1$ case 
for which the potentials 
and the eigenfunctions are
\ba
V^{(-)}_{1}(x)&=&(B^2+A(A+1))\cosech^2{x}-B(2A+1)\cosech {x}\coth {x}\nonumber\\
&+&2\bigg[\frac{(2A+1)}{(2B\cosh x-2A-1)}-\frac{(4B^2-(2A+1)^2)}{(2B\cosh x-2A-1)^2}\bigg]+A^2.
\ea 
and
\be
\psi^{(-)}_{n,1}(x)=N^{(\alpha,\beta)}_{n,1}\frac{(\cosh x-1)^{\frac{(B-A)}{2}}(\cosh x+1)^{-\frac{(B+A)}{2}}}{P^{(-\alpha-1,\beta-1)}_{1}(z)}\hat{P}^{(\alpha,\beta)}_{n+1}(z)
\ee 
respectively. As usual the energy eigenvalues are same as that of the $X_m$ case (\ref{engpt}).
The partner potential corresponding to this case are  
\ba
V^{(+)}_{1}(x)&=&(B^2+A(A-1))\cosech^2{x}-B(2A-1)\cosech {x}\coth {x}\nonumber\\
&+&2\bigg[\frac{(2A-1)}{(2B\cosh x-2A+1)}-\frac{(4B^2-(2A-1)^2)}{(2B\cosh x-2A+1)^2}\bigg]+A^2,
\ea 
and the eigenfunction can be obtained from Eq. (\ref{wfgptpm}). The scattering 
amplitude (\ref{smgpt}) for the $X_1$ case is 
given by  
\be\label{scam1}
s^{(-)}_{1}(k')=2^{-4ik'}\frac{\Gamma(2ik')\Gamma(-A-ik')\Gamma(B-ik'+1/2)(B^2-(ik'-1/2)^2)}{\Gamma(-2ik')\Gamma(-A+ik')\Gamma(B+ik'+1/2)
(B^2-(ik'+1/2)^2)}.
\ee
Following the same procedure as discussed in the last two examples it is 
straight forward tp calculate the integral $I_1(x)$. 
As an illustration, we have calculated it in case $A= 1, B = 3$ and we find
\be\label{intgpt}
I_1(x)=\bigg[\frac{(3+11\cosh x+\cosh 2x)\sech^2 \frac{x}{2}\tanh^5 \frac{x}{2}}{(-2+4\cosh x)}\bigg]
\ee
so that now it is straight forward to obtain the corresponding one parameter
family of strictly isospectral potentials $\hat{V}^{(-)}_1(\lambda, x)$ in case 
$\lambda > 0$ or $\lambda < -1$ and also obtain the corresponding 
eigenfunctions. Further, for $\lambda = 0, -1$ it is straight forward to obtain
the corresponding extended Pursey and Abraham-Moses potentials. In Figs. 3 (a) and 
3(b), we have shown the plots of different RE isospectral potentials 
in case $\lambda \ge 0$ and $\lambda \leq  -1$ respectively. In Fig. 3 (c) we 
have shown the plots of the three isospectral RE potentials $V^{(+)}_1, V^{[P]}_1 
$ and $V^{[AM]}_1$ while in Fig. 3 (d) we have shown the plots of the ground state
eigenfunctions of $\hat{V}^{(-)}(\lambda, x)$ in the case of few positive values of
$\lambda $.

\includegraphics[scale=1.7]{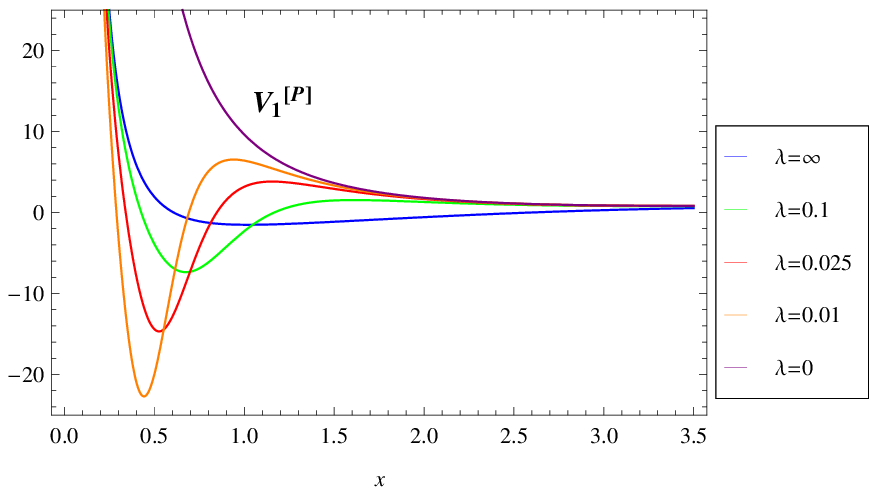}\\
{\bf Fig.3}: {(a) {\it Rationally extended potentials $\hat{V}_1(\lambda,x)$ strictly isospectral to the extended 
GPT potential $V^{(-)}_1(x)$ for $\lambda=0, 0.01,0.025,0.1$ and $\infty$.}\\
\includegraphics[scale=1.7]{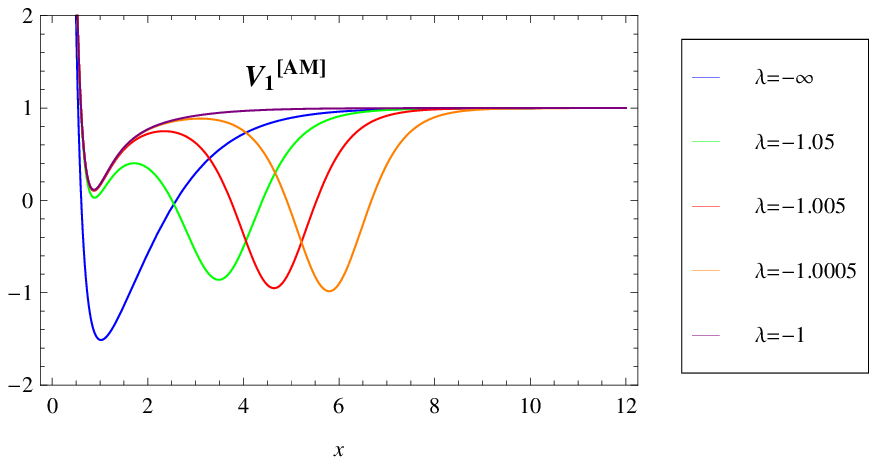}\\
{\bf Fig.3}: {(b) {\it Rationally extended potentials $\hat{V}_1(\lambda,x)$ 
for negative $\lambda=-\infty,-1.05,-1.005,-1.0005$ and $-1$.}\\ 
 \includegraphics[scale=1.6]{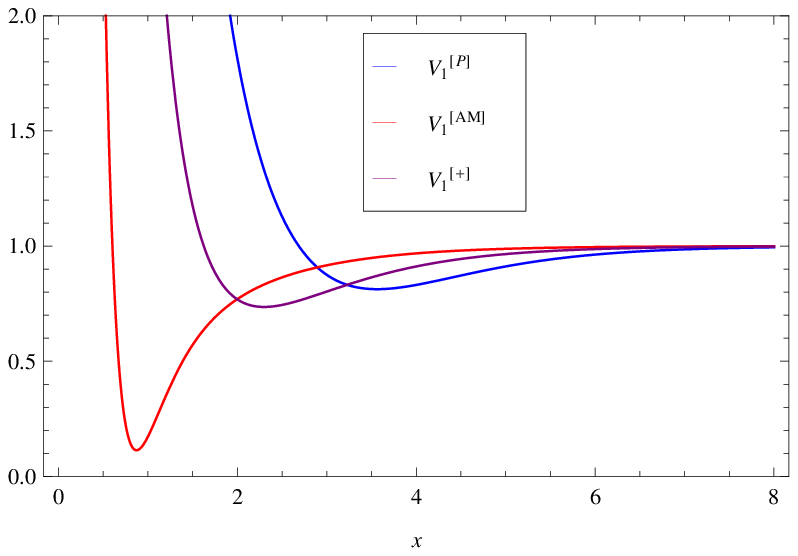}\\
 {\bf Fig.3}: {(c) {\it The extended Pursey potential $V^{[P]}_1(x)$, 
extended AM potential $V^{[AM]}_1(x)$ and the 
extended partner potentials $V^{(+)}_1(x)$.}\\
 \includegraphics[scale=1.6]{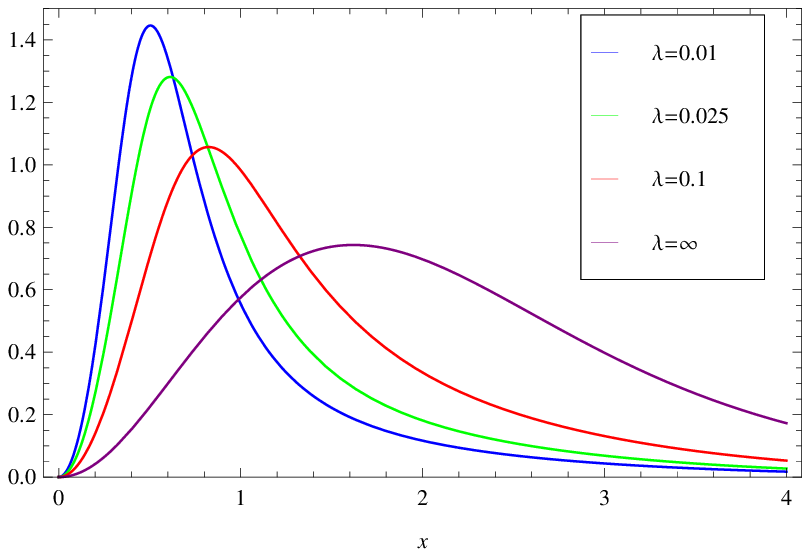}\\
  {\bf Fig.3}: {(d) {\it Normalized ground-state eigenfunctions $\hat{\psi}_{0,1}(\lambda,x )$ for all the potentials 
with positive $\lambda$ shown in Fig. $3(a)$. }\\


The scattering amplitudes $s^{(-)}_1(k')$ (\ref{scam1}) becomes 
\be
s^{(-)}_{1}(k')=2^{-4ik'}\frac{\Gamma(2ik')\Gamma(-1-ik')\Gamma(3-ik'+1/2)(9-(ik'-1/2)^2)}{\Gamma(-2ik')\Gamma(-1+ik')\Gamma(3+ik'+1/2)
(9-(ik'+1/2)^2)}.
\ee
Thus the scattering amplitudes (\ref{3.10}) corresponding to the partner potential, Pursey and AM potentials are given as   
\ba
s^{(+)}_1(k')&=&\bigg(\frac{1+ik'}{1-ik'}\bigg)s^{(-)}_1(k')\nonumber\\
s^{[P]}_1(k')&=&s^{(-)}_1(k')\nonumber
\ea 
and
\be
s^{[AM]}_1(k')=\bigg(\frac{1+ik'}{1-ik'}\bigg)^2 s^{(-)}_1(k')
\ee
respectively.

\section{Conclusions}

In this paper we have considerably extended the list of the exactly solvable 
rationally extended potentials. In particular, by starting from three conventional 
(non-rational) potentials, we have obtained two  
parameter  family of exactly 
solvable strictly isospectral RE potentials by using the formalism of
SUSY Quantum mechanics. Out of these two parameters, one, i.e. $\lambda$ is 
a continuous parameter while $m$ is a discrete parameter ($m = 1, 2, 3,...$)
corresponding to $X_m$ EOPs. We considered three RE potentials namely the RE 
radial oscillator,
RE Scarf-I and the RE generalized P\"oschl-Teller potentials as 
starting potentials and generated one continuous parameter family of strictly
isospectral RE potentials corresponding to these potentials. Further, we have
also obtained the corresponding bound state  
eigenfunctions in terms of the $X_m$ EOPs. Besides, in the case of the RE GPT, 
we have also obtained the scattering amplitude of these potentials. In 
addition, in all these cases we have also obtained the corresponding RE partner
potential $V^{(+)}_m$, the RE Pursey potential $V^{[P]}_m$ and the RE Abraham-Moses
potential $V^{[AM]}_m$ and their bound state eigenfunctions in terms of $X_m$
EOPs. It is worth
pointing out that while these three potentials are strictly isospectral
in the case of the RE radial oscillator and the RE Scarf-I, they are only 
isospectral
(but not strictly isospectral) in the case of the RE generalized 
P\"oschl-Teller potential. 
As an illustration, we have considered the $X_1$ case in detail and shown the 
behavior of these extended potentials as well as the behaviour of the 
corresponding  
normalized ground state wavefunctions for different values (positive as well 
as negative) of $\lambda$. 

Finally, by starting from the ground state of each of these new potentials
and repeating exactly the same procedure as adopted in this paper, in each
case we can further enlarge the number of exactly solvable RE potentials
by obtaining in each case yet another one continuous parameter family of 
strictly isospectral potentials as well as new RE Pursey and RE Abraham-Moses
potentials.  

{\bf Acknowledgments}\\
BPM acknowledges the support 
from MATRIX project (Grant No. MTR/2018/000611), SERB, DST Govt. of India.  AK is grateful to Indian National Science Academy (INSA) for 
awarding INSA senior scientist position at Savitribai Phule Pune University.


\begin{thebibliography}{99}
\bibitem{cks}  F. Cooper, A. Khare, U. Sukhatme {\it  Phys. Rep.  } \textbf{251} (1995) 267; {\it  "SUSY in Quantum Mechanics"  } World Scientific (2001).
\bibitem{ks}  A. Khare and U. P. Sukhatme {\it J. Phys. A: Math. Gen.22(1989) 2847 }
\bibitem{invs1} Z. S. Agronorich and V. A. Marchenko, {\it "The Inverse Problem of Scattering Theory"}  Gordon and Breach, New York (1963).
\bibitem{invs2} K. Chandan and P. C. Sabatier, {\it "Inverse Problem in Quantum Scattering Theory"} Springer, New York (1977).
\bibitem{baye}  D Baye, {\it Phys. Rev. Lett.} \textbf{58} (1987) 2738.
\bibitem{sol1}  G. L. Lamb, D Baye, {\it "Elements of Soliton Theory"} Springer, New York (1980).
\bibitem{sol2}  P. G. Drazin and R. S. Johson, {\it "Solitons: An Introduction"} Cambridge University Press (1989).
\bibitem{nieto} M.M. Nieto, {\it Phys. Lett. B} \textbf{145} (1984) 208.
\bibitem{amado} R. D. Amado, {\it Phys. Rev. A} \textbf{37} (1988) 2277.
\bibitem{pursey} D. L. Pursey, {\it Phys. Rev. D} \textbf{33 1048} (1986) 2267.
\bibitem{cv} C. V. Sukumar, {\it J. Phys. A: Math. Gen. } \textbf{21} (1988) L455; {\it J. Phys. A: Math. Gen. } \textbf{18} (1985) 2937. 
\bibitem{amp} P. B. Abraham and H.E. Moses, {\it Phys. Rev. A} \textbf{22} (1980) 1333.
\bibitem{darboux} G. Darboux, {\it C. R. Acad. Sci. (Paris)} \textbf{94} (1882) 1456.
\bibitem{akus} A. Khare and U. Sukhatme, {\it J. Phys. A: Math. Gen.} \textbf{22} (1989) 2847.
\bibitem{gele} I. M. Gelfand and B. M. Levitan, {\it Am. Math. Soc. Transl.} \textbf{1} (1951) 253.
\bibitem{eopm1} D. Gomez-Ullate, N. Kamran and R. Milson, {\it J. Math. Anal.Appl.} \textbf{359} (2009) 352.  
\bibitem{eopm2} D. Gomez-Ullate, N. Kamran and R. Milson, {\it J. Phys. A} \textbf{43} (2010) 434016.
\bibitem{eopm3} D. Gomez-Ullate, N. Kamran and R. Milson, {\it Contemp. Math.} \textbf{563} (2012) 51.
\bibitem{que}  C. Quesne, {\it J.Phys.A} \textbf{41} (2008) 392001.
\bibitem{bqr}  B. Bagchi, C. Quesne and R. Roychoudhary, 
{\it Pramana J. Phys.} \textbf{73}(2009) 337, C. Quesne, SIGMA {\bf 5} (2009)
84.
\bibitem{os}  S. Odake and R. Sasaki, {\it Phys. Lett. B}, \textbf{684} 
(2010) 173; ibid {\bf 679} (2009) 414. {\it J. Math. Phys}, \textbf{51}, 053513 (2010).
\bibitem{hos} C-L. Ho, S ODAKE and R Sasaki, {\it SIGMA} \textbf{7} (2011) 107. 
\bibitem{hs} C-L. Ho and R Sasaki, {\it ISRN Math. Phys.} (2012) 920475.
\bibitem{que8}  C. Quesne, {\it Int. J. Mod. Phys. A} \textbf{26} (2011) 5337.
\bibitem{op1} Y. Grandati, {\it J. Math. Phys.} \textbf{52} 103505 (2011).
\bibitem{op2} Y. Grandati, {\it Ann. Phys.} \textbf{326} 2074 (2011); \textbf{327} 2411 (2012); \textbf{327} 185 (2012).
\bibitem{op3} C. Quesne, {\it SIGMA} \textbf{8} 080 (2012).
\bibitem{midya}  B Midya and B Roy, {\it J. Phys. A: Mathematical and Theoretical} \textbf{46 (17)} 175201 (2013).
\bibitem{rkyd} R. K. Yadav, B. P. Mandal and A. Khare, {\it Acta Polytechnica} \textbf{57(6)} (2017) 477.
\bibitem{midyapd}  B Midya and B Roy, {\it Phys. Lett. A} \textbf{373 (45)} (2009) 4117.
\bibitem{midya1}  B Midya, B Roy and T. Tanaka, {\it J. Phys. A} \textbf{ 45} (2012) 205303.
\bibitem{clh11}  C.-L Ho, {\it Ann. Phys.} \textbf{326} (2011) 797.
\bibitem{dr11}  D. Dutta and P. Roy, {\it J. Math. Phys.}\textbf{52} (2011) 032104.
\bibitem{scatt1}  R. K. Yadav, A. Khare and B. P. Mandal, {\it Ann. Phys.} \textbf {331} (2013) 313.
\bibitem{scatt}  R. K. Yadav, A. Khare and B. P. Mandal, {\it Phys. Lett. B} \textbf {723} 433 (2013); {\it Phys. Lett. A} \textbf {379} (2015) 67.
\bibitem{scatt4}  C. L. Ho, J. C. Lee and R. Sasaki, {\it Annals of Physics} \textbf{343} (2014) 115.
\bibitem{n16}  R. K. Yadav, N. Kumari, A. Khare and B. P. Mandal, {\it Ann. Phys.} \textbf{359} (2015) 46.
\bibitem{nk16}  N. Kumari, R. K. Yadav, A. Khare, B. Bagchi and B. P. Mandal, {\it Ann. Phys.}\textbf{373} (2016) 163.
\bibitem{ramos} A. Ramos et al., {\it Ann. Phys.}\textbf{382} (2017) 143.
\bibitem{para}   R. K. Yadav, A. Khare, N. Kumari, B. Bagchi and B. P. Mandal, {\it J. Math. Phys.}\textbf{57} (2016) 062106.
\bibitem{nk17}  N. Kumari, R. K. Yadav, A. Khare and B. P. Mandal, {\it Ann. Phys.}\textbf{385} (2017) 57.
\bibitem{nk18}  N. Kumari, R. K. Yadav, A. Khare and B. P. Mandal, {\it J. Math. Phys.}\textbf{59} (2018) 062103-1.
\bibitem{bbp}  B. Basu-Mallick, B. P. Mandal and P. Roy, {\it Ann. Phys.}\textbf{380} (2017) 206.
\bibitem{rkmany}  R. K. Yadav, A. Khare, N. Kumari and B. P. Mandal, {\it Ann. Phys.}\textbf{400} (2019) 189.


 


\end{thebibliography}
\end{document}